\documentclass[aps,prl,twocolumn,showpacs,superscriptaddress]{revtex4}

\usepackage{graphicx}         % needed for figures
\usepackage{bm}               % for math
\usepackage{amssymb}          % for math
\usepackage{amsmath}          % for aligned

\def \eV          {\rm eV}
\def \pc          {\rm pc}
\def \kpc         {\rm kpc}
\def \Mpc         {\rm Mpc}
\def \msun        {M_\odot}
\def \psiDM       {$\psi$DM}

\begin{document}

\title{Understanding the Core-Halo Relation of Quantum Wave Dark Matter, {\psiDM}, from 3D Simulations}

%authors and affiliations
\author{Hsi-Yu Schive}
%\email{hyschive@ntu.edu.tw}
\affiliation{Department of Physics, National Taiwan University, Taipei 10617, Taiwan}

\author{Ming-Hsuan Liao}
\affiliation{Department of Physics, National Taiwan University, Taipei 10617, Taiwan}

\author{Tak-Pong Woo}
\affiliation{Department of Physics, National Taiwan University, Taipei 10617, Taiwan}

\author{Shing-Kwong Wong}
\affiliation{Department of Physics, National Taiwan University, Taipei 10617, Taiwan}

\author{Tzihong Chiueh}
\email{chiuehth@phys.ntu.edu.tw}
\affiliation{Department of Physics, National Taiwan University, Taipei 10617, Taiwan}
\affiliation{National Center for Theoretical Sciences, National Taiwan University, Taipei 10617, Taiwan}

\author{Tom Broadhurst}
\affiliation{Department of Theoretical Physics, University of the Basque Country UPV/EHU, E-48080 Bilbao, Spain}
\affiliation{Ikerbasque, Basque Foundation for Science, E-48011 Bilbao, Spain}

\author{W-Y. Pauchy Hwang}
\affiliation{Department of Physics, National Taiwan University, Taipei 10617, Taiwan}
\affiliation{National Center for Theoretical Sciences, National Taiwan University, Taipei 10617, Taiwan}

%\date{\today}

% abstract
% ----------------------------------------------
\begin{abstract}

We examine the nonlinear structure of gravitationally collapsed objects
that form in our simulations of wavelike cold dark matter ({\psiDM}),
described by the Schr\"{o}dinger-Poisson (SP) equation with a particle mass $\sim 10^{-22}~\eV$. A
distinct gravitationally self-bound solitonic core is found at the center of
every halo, with a profile quite different from cores modeled in the warm or
self-interacting dark matter scenarios. Furthermore, we show that each
solitonic core is surrounded by an extended halo composed of large fluctuating
dark matter granules which modulate the halo density on a scale comparable to the
diameter of the solitonic core. The scaling symmetry of the SP equation and the
uncertainty principle tightly relate the core mass to the halo specific
energy, which, in the context of cosmological structure formation,
leads to a simple scaling between core mass ($M_c$)
and halo mass ($M_h$), $M_c \propto a^{-1/2} M_h^{1/3}$,
where $a$ is the cosmic scale factor. We verify this scaling relation by
(i) examining the internal structure of a statistical sample of
virialized halos that form in our 3D cosmological simulations, and by
(ii) merging multiple solitons to create individual virialized objects.
Sufficient simulation resolution is achieved by adaptive mesh refinement
and graphic processing units acceleration. From this scaling relation, present dwarf
satellite galaxies are predicted to have kpc sized cores and a minimum mass
of $\sim 10^8~\msun$,
capable of solving the small-scale controversies in the cold dark matter model.
Moreover, galaxies of $2\times10^{12}~\msun$ at $z=8$
should have massive solitonic cores of $\sim 2\times10^9~\msun$
within $\sim 60~\pc$. Such cores can provide a favorable local
environment for funneling the gas that leads to the prompt formation of early
stellar spheroids and quasars.

\end{abstract}

\maketitle

% introduction 
% ----------------------------------------------
Accumulating evidences suggest that the Universe contains $\sim 26\%$
dark matter \cite{Planck2014} which interacts primarily through self-gravity.
Dark matter comprising very light bosons with a mass
$m_{\psi} \sim 10^{-22}~\eV$ has been recognized as a viable means of
suppressing low mass galaxies and providing cored profiles in dark matter
dominated galaxies \cite{Hu2000,Peebles2000}.
Interestingly, this boson mass scale can naturally arise in a non-QCD
axion model \cite{Chiueh2014}, lending support for the very light boson.
The relative deficiency of the
observed number of low-mass galaxies is a major problem for standard cold
dark matter (CDM) \cite{Kauffmann1993,Klypin1999,Moore1999}, for which a
steeply rising mass function is predicted \cite{Diemand2007}. Furthermore,
the dwarf spheroidal galaxies
\cite{Moore1994,Flores1994,Kleyna2003,Goerdt2006,Gilmore2007,Battaglia2008,
WP2011,AE2012,Cole2012,JR2012,Penarrubia2012,AAE2013}
and low surface brightness galaxies \cite{deBlok2001,deBlok2002} are generally
inferred to have large flat cores of dark matter, at odds with the singular
cores required by standard CDM \cite{DC1991,NFW}.
Complicated baryonic physics such as supernova feedback is required
to solve both issues in the CDM paradigm \cite{Navarro1996,Read2005,Read2006,
Penarrubia2010,Zolotov2012,Pontzen2012,Teyssier2013,Brooks2014,Pontzen2014,Kauffmann2014}.
   
Extremely light bosonic dark matter can be assumed to be non-thermally generated
and described by a single coherent
wave function \cite{Hu2000,Goodman2000,Bohmer2007,Sikivie2009,WC2009}, which
we term {\psiDM}. Here solutions to both the
missing-satellite and cusp-core problems arise from the uncertainty
principle, leading to an effective quantum-mechanical stress tensor that
suppresses small-scale structures below a Jeans scale. The Jeans scale
evolves with the cosmic time slowly as $a^{-1/4}$, where $a$ is the cosmic
scale factor \cite{Hu2000,WC2009}, thereby yielding a sharp break in the
linear mass power spectrum. This expected behaviour has recently been
demonstrated with the first cosmological simulations at sufficiently
high resolution, capable of resolving the smallest galaxy halos forming in
this context \cite{Schive2014}.

Warm dark matter (WDM) is also capable of suppressing small-scale linear
power by free streaming \cite{Bode2001}, but it suffers from the
\emph{Catch 22} problem \cite{Maccio2012}, where the light particle mass
required for creating a sufficiently large core ($\sim 1~\kpc$) would prevent
the formation of dwarf galaxies in the first place. Collisional
CDM does somewhat better in producing cores consistent with observations,
but it cannot suppress the number of dwarf galaxies
\cite{SS2000,Rocha2013}. For these reasons, {\psiDM} and scalar-field dark
matter composed of extremely light particles have recently begun to attract
attention as a viable contender for the long-sought dark matter
(e.g., \cite{SM2011,Chavanis2011,Lora2012,RM2012,Rindler2014,Huang2014,KR2014,MS2014,Schive2014}).
     
Cosmic structures at high redshifts provide stringent tests for all
alternative dark matter models attempting to solve the small-scale issues of
CDM in the Local Group. For WDM a tension arises when requiring the
relatively large cores of dwarf spheroidal galaxies without violating the
small scale power constrained by the Lyman-$\alpha$ forest
\cite{Strigari2006,Maccio2012,Viel2013,Schneider2014}. For {\psiDM} this
problem may be less severe due to the sharper small-scale break in its linear
power spectrum as compared to WDM \cite{Hu2000,MS2014}. The power spectrum is
marginally consistent with the Lyman-$\alpha$ forest observations, while
adding a small amount of CDM component ($\sim 10\%$) can certainly further
relieve the tension \cite{MS2014}. High-$z$ number counts provide another
constraint for galaxies at $6 \le z \le 8$ \cite{Schultz2014}.
We notice that the {\psiDM} power spectrum starts to deviate from CDM at
$k \sim 7~h~\Mpc^{-1}$ \cite{Schive2014}, corresponding to a halo mass of
$\sim 5\times10^9~\msun$. Above this mass scale the {\psiDM} galaxy number
density should be close to CDM, and therefore consistent with the observational
constraint \cite{Schultz2014,Bozek2014}. Larger {\psiDM} simulations with the addition
of baryons will be invaluable for supporting these arguments
and testing with the forthcoming observations such as JWST \cite{Gardner2006} and AdvACT \cite{Calabrese2014}.

\begin{figure}[t]
\centering
\includegraphics[width=8.6cm]{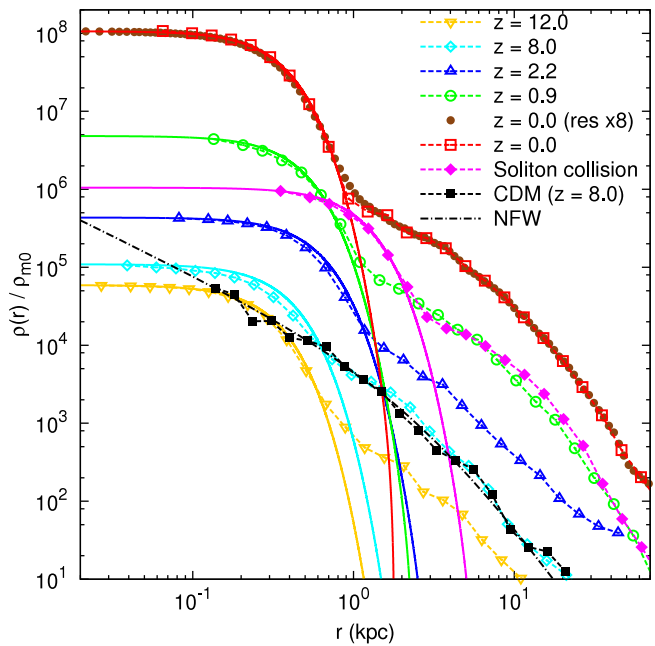}
\caption{Density profiles of {\psiDM} halos.
Dashed lines with various opened symbols show five examples at
different redshifts between $12 \ge z \ge 0$. The DM density is normalized to
the cosmic background density. A distinct core forms in every halo as a
gravitationally self-bound object, satisfying the redshift-dependent soliton
solution (solid lines) upon proper $\lambda$ scaling.
As a convergence test, filled circles show the same $z=0$ halo (the most
massive one) but with eight times higher resolution.
Filled diamonds show an
example from the soliton collision simulations arbitrarily renormalized
to the comoving coordinates at $z=0$.
The same $z=8$ halo in a CDM simulation (filled squares) fit by an NFW
profile (dot-dashed line) is also shown for comparison.}
\label{fig:CoreProfile__LSS}
\end{figure}

Previous theoretical work on {\psiDM} halos mainly focused on two aspects:
(i) a stationary soliton profile with or without self-interaction (e.g.,
\cite{Goodman2000,Bohmer2007,Lora2012}), or (ii) a Navarro-Frenk-White (NFW)
profile \cite{NFW} with its inner cusp replaced by a flat core
(e.g., \cite{Hu2000,MS2014}). In either case, the detailed connection between
cores and halos in the fully nonlinear regime has not been addressed. This 
question can be best answered by simulations. The first attempt of
three-dimensional simulations of the {\psiDM} structure formation has come
to light only a few years ago \cite{WC2009}, revealing complex interference
fringes and a halo profile similar to NFW. This work however did not have
sufficient spatial resolution for resolving the innermost cores. More
recently, Schive et al. \cite{Schive2014} made a great leap forward in the
{\psiDM} simulations by taking advantage of an
adaptive mesh refinement (AMR) scheme powered by graphic processing units
(GPU) acceleration \cite{GAMER2010}. A prominent solitonic core is found in
every halo, appearing as a self-bound mass clump superposed on the NFW
profile (see Fig. \ref{fig:CoreProfile__LSS}). This surprising core
configuration is apparently different from the linear prediction of
{\psiDM} \cite{MS2014}, WDM \cite{Maccio2012}, and collisional
dark matter \cite{Rocha2013}, in all of which a constant-density core 
is introduced truncating the otherwise cuspy NFW profile.
Using the stellar phase-space distribution of
the Fornax dwarf spheroidal galaxy, the soliton profile is found to
be consistent with observations assuming 
$m_{\psi}=(8.0^{+1.8}_{-2.0})\times 10^{-23}~\eV$.
Furthermore, this work demonstrates that {\psiDM} can clear the
\emph{Catch 22} problem facing WDM.

In this Letter, we examine the relationship between the solitonic core and
the host halo, which we quantify statistically with
simulations. We demonstrate that the solitonic core and
the halo always coexist in a relaxed, self-bound system of {\psiDM}.
The core mass is tightly related to the halo specific energy, which, for
cosmological structure formation, leads to a simple
redshift-dependent core-halo mass relation.

Wave mechanics of {\psiDM} is governed by the Schr\"{o}dinger-Poisson
(SP) equation \cite{Seidel1990,Widrow1993}. In an expanding universe, the equation can
be written in the comoving coordinates as
\begin{equation}
\left[i\frac{\partial}{\partial\tau}+\frac{\nabla^2}{2}-aV\right]\psi=0
\label{eq:Schrodinger}
\end{equation}
and
\begin{equation}
\nabla^2 V= 4\pi(|\psi|^2-1),
\label{eq:Poisson}
\end{equation}
where the comoving length $\bm x$ is normalized to
$(\frac{3}{8\pi}H_0^2\Omega_{m0})^{-1/4}(m_{\psi}/\hbar)^{-1/2}$,
the time normalized to
$d\tau\equiv (\frac{3}{8\pi}H_0^2\Omega_{m0})^{1/2}a^{-2}dt$,
and the wave function $\psi$ normalized to $(\rho_{m0}/m_{\psi})^{1/2}$. Here
$H_0$, $\Omega_{m0}$ and $\rho_{m0}$ are the present Hubble parameter,
matter density parameter and background mass density, respectively, and
$V$ the gravitational potential. An important feature of the
SP equation is its scaling symmetry \cite{Seidel1990,Guzman2006}.
It can be easily seen that when $|\psi|^2 \gg 1$ and $a=const.$,
the SP equation remains unchanged under the transformation
$(\tau,{\bm x},\psi,V)\to (\lambda^{-2}\tau,\lambda^{-1}{\bm x},\lambda^2\psi,\lambda^2 V)$
for arbitrary $\lambda$.
Having very high densities and forming in a short time compared with the
Hubble time, all solitonic cores hence conform to this $\lambda$ scaling
to a high accuracy. The relevant physical quantities scale as
$(x_c, \rho_c, M_c, E_c) \to
(\lambda^{-1} x_c,\lambda^4 \rho_c,\lambda M_c,\lambda^3 E_c)$,
where $x_c$, $\rho_c$, $M_c$ and $E_c$ are the core radius, density, mass
and energy, respectively. The soliton density profile can be well fit
by \cite{Schive2014}
\begin{equation}
\rho_c(x) = \frac{1.9~a^{-1}(m_\psi/10^{-23}~\eV)^{-2}(x_c/\kpc)^{-4}}{[1+9.1\times10^{-2}(x/x_c)^2]^8}~\msun\pc^{-3},
\label{eq:SolitonFit}
\end{equation}
accurate to $2\%$ in the range $0 \le x \lesssim 3\,x_c$.
Here we define $x_c$ as the radius at which the density drops to one-half
its peak value, and $M_c$ as the enclosed mass within $x_c$.
Note that $M(x \le 3\,x_c)$ makes up about $95\%$ of the total soliton
mass, and the half-mass radius is $\sim 1.45~x_c$.

To address the core-halo configuration, we conduct three structure formation
simulations of different realizations with a spatial resolution up to 60 pc
in a 2 Mpc comoving box. These runs begin at the matter-radiation equality
around $z=3,200$ and end at $z=0$.
Note that the small simulation box will affect the statistical
properties of halos such as the mass function \cite{Reed2007}, but
should have a small impact on the core-halo relation addressed in this Letter
which mainly relies on the virialization of each individual halo and is 
insensitive to the initial power spectrum. We
demonstrate this point by tracing several halos in a 20 Mpc box with the same
spatial resolution as in the 2 Mpc simulations.
Another simulation with a 40 Mpc
box is conducted from $z=3,200$ to $z=8$ for probing the high-redshift
galaxies. Our results verify that halos at different
redshifts all contain self-similar solitonic cores. Density granules of about
the same size as the solitonic core are apparent throughout the halos (see
Fig. 2 in Ref. \cite{Schive2014} for an illustration): an important
feature for the core-halo connection and will be explained later. The soliton
profile is redshift-dependent. To see this, note that as long as $a$ can be
regarded as a constant, the SP equation can be rewritten into a
redshift-independent form by introducing a set of rescaled variables:
$(\tau', {\bm x'} ,\psi', V') \equiv (a^{1/2} \tau, a^{1/4} {\bm x}, \psi, a^{1/2} V)$.
It follows that the soliton radius in the
comoving (unprimed) coordinates scales as $a^{-1/4}$ for a fixed peak core
density. Figure \ref{fig:CoreProfile__LSS} shows the density profiles of
typical halos in the simulations at five different epochs,
$z=12.0,~8.0,~2.2,~0.9~\rm{and}~0.0$, in the unprimed coordinates. The agreements
of the simulation data to both the $\lambda$ and $a$ scalings are excellent.

% Fig. 1
% ----------------------------------------------

\begin{figure}[t]
\centering
\includegraphics[width=8.6cm]{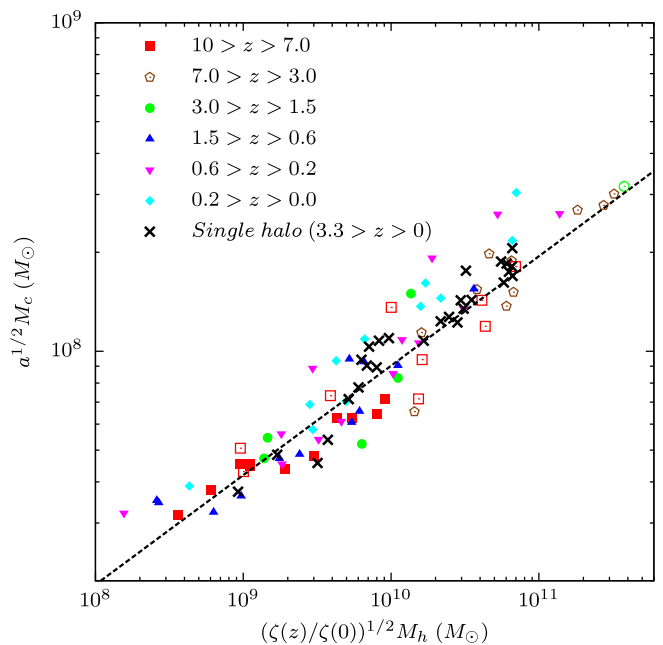}
\caption{Core-halo mass relation. Different filled symbols show halos at
different epochs in the 2 and 40 Mpc simulations, and open symbols represent
the 20 Mpc simulation. Crosses trace the evolution of a single halo.
Dashed line shows the analytical prediction given by
Eq. (\ref{eq:CoreM_vs_HaloM}) (see text for details).}
\label{fig:CoreM_vs_HaloM__LSS}
\end{figure}

A question naturally arises concerning the relation between solitonic cores
and their host halos. Aided by our structure formation simulations, we find
all collapsed objects approximately follow a redshift-dependent core-halo
mass relation,
\begin{equation}
M_c \propto a^{-1/2} M_h^{1/3}.
\label{eq:SolitonHaloScaling}
\end{equation}
The halo virial mass is defined as
$M_h \equiv (4\pi x_{vir}^3 / 3) \zeta(z) \rho_{m0}$,
where $x_{vir}$ is the comoving virial radius and
$\zeta(z) \equiv (18\pi^2 + 82(\Omega_m(z)-1) - 39(\Omega_m(z)-1)^2)/\Omega_m(z) \sim 350~(180)$ at $z=0~(z\geq1)$
\cite{BN1998}. Note that this definition
of virial mass is the same as that for CDM. This is because once an object
exceeds the Jeans mass on its way to collapse, the dynamics is almost
identical to the cold collapse, for which the Eikonal approximation of wave
dynamics to particle dynamics holds until virialization takes place. Figure
\ref{fig:CoreM_vs_HaloM__LSS} shows this scaling relation over
three orders of magnitude in halo mass from $10^8$ to $5 \times 10^{11}~\msun$.
We demonstrate the redshift evolution by showing
coalescence of the core-halo mass relations of halos at different
epochs between $10>z>0$ as well as the evolutionary trajectory of a single
halo.
Note that low-redshift, massive halos in the 2 Mpc runs show a relatively larger
scatter, which could be due to the small box effect, while massive halos in
the 20 Mpc run do converge to our analytical prediction. In
all cases the deviation of the core mass from Eq. (\ref{eq:SolitonHaloScaling})
is less than a factor of two.
Also note that the halos in the simulations with a mass several
times $10^8~\msun$ are found to be dominated by the central solitons, a key for
estimating the minimum halo mass as will be discussed later.

% Fig. 2
% ----------------------------------------------

\begin{figure}[t]
\centering
\includegraphics[width=8.6cm]{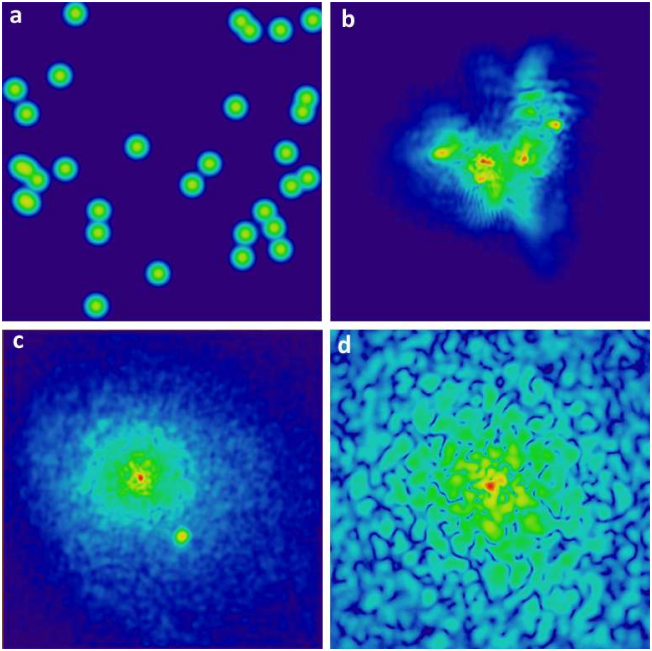}
\caption{Snapshots of a soliton collision simulation. Panels (a)-(c) show the
projected density distribution at the initial and intermediate stages, and
panel (d) shows a close-up of the conspicuous solitonic core at the final stage.
Fluctuating density granules resulting from the quantum wave 
interference appear everywhere and have a size similar to the central soliton.}
\label{fig:Dens2D__SolitonMerger}
\end{figure}

To understand this core-halo mass relation, we further conduct a set of
controlled numerical experiments, where multiple solitons are initially
placed randomly with zero velocity and start to merge until the systems relax.
Solitons are chosen as a convenient initial condition for their stability. 
Here we assume $a=const.$ and zero background density. We would like to
know whether the core-halo configuration still persists in a different setting from
cosmological structure formation, and if so, we want to ascertain what factors
determine the soliton scale among the infinite number of self-similar 
solutions. Intuitively, one expects that the final relaxed state
should lose the memory of its initial configuration and thus depends only on
the globally conserved quantities, namely, the total mass $M$ and energy $E$
(assuming there is no net angular momentum). We conduct 29 runs in total
with different initial conditions of various $M$ and $E$. For
the same $M$ and $E$, we repeat runs with different realizations, including
different initial soliton numbers ranging from 4 to 128, different soliton
sizes and initial positions. Figure \ref{fig:Dens2D__SolitonMerger} shows one
example of the soliton collision simulations. The AMR scheme
is again adopted in order to achieve sufficient resolution everywhere; in
particular, we ensure that every soliton is well resolved with at least
$\sim 10^4$ cells and verify that $M$ and $E$ remain
conserved with at most a few percent error in all simulations.

% Fig. 3
% ----------------------------------------------

The resulting relaxed structures that form in these soliton collision
experiments are always found to consist of a halo and a solitonic core (see
Fig. \ref{fig:CoreProfile__LSS} and panel (d) of
Fig. \ref{fig:Dens2D__SolitonMerger}), similar to the results of cosmological
simulations. The core profiles satisfy the $\lambda$ scaling and the halo
profiles are close to NFW. This result establishes that the core-halo
configuration is a generic structure of {\psiDM} in
virialized gravitational equilibrium.

\begin{figure}[t]
\centering
\includegraphics[width=8.6cm]{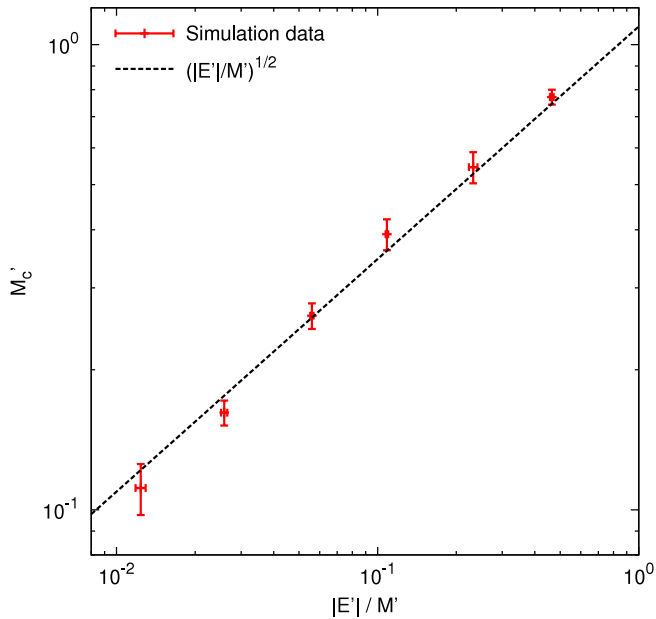}
\caption{Scaling relation between core mass and system specific energy
in the soliton collision experiments. Error bars represent the
root-mean-square scatter of different realizations at a
given specific energy bin as well as the fluctuation in different snapshots of each run.
Note that the redshift dependence has been
absorbed into the rescaled mass $M'$ and energy $E'$ (see text for details)}.
\label{fig:SolitonEMScaling__SolitonMerger}
\end{figure}

More importantly, as shown in Fig. \ref{fig:SolitonEMScaling__SolitonMerger},
the core mass follows the relation
\begin{equation}
M'_c = \alpha (|E'|/M')^{1/2}.
\label{eq:SolitonEMScaling__SolitonMerger}
\end{equation}
Here the total kinetic energy, potential energy and mass are defined in the
primed (redshift-independent) coordinates as
$E'_k \equiv \frac{1}{2}\int |\nabla'\psi'|^2 d^3 x'$,
$E'_p\equiv \frac{1}{2}\int |\psi'|^2 V' d^3x'$,
$M'\equiv\int |\psi'|^2 d^3 x'$, and $\alpha$ is a dimensionless constant
close to unity. The physical foundation of this relation can be appreciated
as follows. The RHS represents the halo velocity dispersion, $\sigma'_h$, and
on the LHS the $\lambda$ scaling demands that $M'_c \sim {x'}_c^{-1}$, the
inverse soliton size. Accordingly,
Eq. (\ref{eq:SolitonEMScaling__SolitonMerger}) relates
the soliton size to the halo velocity dispersion through the uncertainty
principle, where $x'_c \sigma'_h \sim 1$. This result is non-trivial in that
the uncertainty principle is originally a local relation, but here it is found
to hold non-locally, relating a core (local) property to a halo (global)
property. The non-local uncertainty principle reveals itself in panel (d)
of Fig. \ref{fig:Dens2D__SolitonMerger}. The inverse halo velocity dispersion
is manifested by the size of halo density granules, and the fact that the halo
granule size is close to the soliton size provides another perspective to view
the finding of Eq. (\ref{eq:SolitonEMScaling__SolitonMerger}). 
Eigenmode decomposition of the core-halo system can help our understanding
of the detailed physics underlying this quantum ``thermalization'',
and it will be presented in a separate work (Wong et al., in preparation).

% Fig. 4
% ----------------------------------------------

We are now in a position to understand the physical meaning of the empirical
Eq. (\ref{eq:SolitonHaloScaling}). In the structure formation simulations,
we verify that halos at different redshifts all conform to
Eq. (\ref{eq:SolitonEMScaling__SolitonMerger}) by taking $E'$ and $M'$ as
the rescaled halo energy ($E'_h$) and virial mass ($M'_h$). Adopting the
virial condition in the spherical collapse model $|E'_h|=|E'_p|/2 \sim 3M_h^{'2}/10x'_{vir}$ and retrieving
the redshift dependence then give $M_c = \alpha(3M_h/10x_{vir})^{1/2}a^{-1/2}$.
Finally, solving $x_{vir}$ as a function of $M_h$ using the definition of
virial mass given immediately after Eq. (\ref{eq:SolitonHaloScaling}) yields
the expected core-halo mass relation
\begin{equation}
M_c = \frac{1}{4} a^{-1/2} \left( \frac{\zeta(z)}{\zeta(0)} \right)^{1/6}
      \left( \frac{M_h}{M_{min,0}} \right)^{1/3} M_{min,0},
\label{eq:CoreM_vs_HaloM}
\end{equation}
where
$M_{min,0} = 375^{-1/4} 32\pi  \zeta(0)^{1/4} \rho_{m0} (H_0 m_{\psi} / \hbar)^{-3/2} \Omega_{m0}^{-3/4}$
$\sim 4.4\times10^7 m_{22}^{-3/2}~\msun$. Here $m_{22} \equiv m_{\psi}/10^{-22}~\eV$
and we have taken $\alpha=1$ and typical values for the cosmological parameters.
Eq. (\ref{eq:CoreM_vs_HaloM}) is consistent with
Eq. (\ref{eq:SolitonHaloScaling}) apart from an additional
slowly varying factor $\zeta(z)^{1/6}$. 
The physical core radius, $r_c=a x_c$, is inversely proportional to $M_c$
and can be expressed as
\begin{equation}
r_c = 1.6~m_{22}^{-1}a^{1/2} \left( \frac{\zeta(z)}{\zeta(0)} \right)^{-1/6}
                             \left( \frac{M_h}{10^9~\msun} \right)^{-1/3}~\kpc.
\label{eq:CoreR_vs_HaloM}
\end{equation}

The smallest halo should be close to a single isolated soliton, with a wide core
and a steeper outer gradient. Our definition of core mass, $M(r\le r_c)$, makes
up about 25\% of the total soliton mass. Thus by taking $M_c = M_h/4$ in
Eq. (\ref{eq:CoreM_vs_HaloM}) we readily obtain a minimum halo mass
$M_{min}(z) = a^{-3/4} (\zeta(z)/\zeta(0))^{1/4} M_{min,0} \sim 3\times10^8~\msun$
at $z=8$ for $m_{22}=0.8$, consistent with Fig. \ref{fig:CoreM_vs_HaloM__LSS}
and the theoretical prediction \cite{Bozek2014}.

Finally, we conclude this Letter by a conjecture regarding the possible
consequences of the early formation of the dense solitonic cores.
A present-day galaxy with a typical halo mass of $2\times10^{12}~\msun$ will
have $M_c \sim 5\times10^8~\msun$ and $r_c \sim 160~\pc$.
For a high-redshift galaxy with the same halo mass, its core mass and
gravitational acceleration near the core, $M_c/r_c^2$, will be enhanced
by a factor of $a^{-1/2}$ and $a^{-3/2}$, respectively. This much greater
gravitational force may quickly attract a large amount of gas into a small
central region, thereby creating an ultra-dense gas favorable for
major starbursts and formation of supermassive black holes. For example,
a galaxy of $2\times10^{12}~\msun$ forming at
$z=8$ has a core mass $\sim 2\times10^9~\msun$ in $\sim 60 ~\pc$ radius and it
captures at least $4\times10^8~\msun$ gas if the baryon fraction at the core
is the same as or above the cosmic mean. If furthermore the gas temperature
maintains near the Lyman-$\alpha$ onset temperature, $10~\eV$, this radius
is only a factor of two greater than the $30~\pc$ thermal Jeans length of the
gas. Such a solitonic core can certainly help the prompt formation of quasars
appearing as early as $z=7$ \cite{Mortlock2011}.

% acknowledgement
% ----------------------------------------------
This work is supported in part by the National Science Council of Taiwan
under the grants NSC100-2112-M-002-018-MY3 and NSC99-2112-M-002-009-MY3.

% references
% ----------------------------------------------
% short names of references
\newcommand {\apjl}  {Astrophys. J. Lett.}
\newcommand {\mnras} {Mon. Not. R. Astron. Soc.}
\newcommand {\na}    {New Astron.}

\bibliography{Reference}

\end{document}